\documentclass[preprint,12pt]{elsarticle}

\usepackage{graphicx}
\usepackage{epstopdf} 
\usepackage{amssymb}
\usepackage{color}
\biboptions{numbers}
\journal{Computer Physics Communications}

\begin{document}
\bibliographystyle{elsarticle-num}

\title{{Simplified numerical model for clarifying scaling behavior in the intermediate dispersion regime in homogeneous porous media}}

\author[ciemat]{B.Ph.~van Milligen\corref{cor1}}
\ead{boudewijn.vanmilligen@ciemat.es}
\author[eku]{P.D. Bons}
\ead{paul.bons@uni-tuebingen.de}
 \address[ciemat]{National Fusion Laboratory, CIEMAT, Avda.~Complutense 40, 28040 Madrid, Spain}
 \address[eku]{Eberhard Karls University, Department of Geosciences, Wilhelmstrasse 56, 72074 T\"ubingen, Germany}
\cortext[cor1]{Corresponding author}

\begin{abstract}
The dispersion of solute in porous media shows a non-linear increase in the transition from diffusion to advection dominated dispersion as the flow velocity is raised. 
In the past, the behavior in this intermediate regime has been explained with a variety of models. 
{We present and use a simplified numerical model which does not contain any turbulence, Taylor dispersion, or fractality.
With it, we show that the non-linearity in the intermediate regime nevertheless occurs.
Furthermore,} we show that that the intermediate regime can be regarded as a phase transition between random, diffusive transport at low flow velocity and ordered transport controlled by the geometry of the pore space at high flow velocities. 
This phase transition explains the first-order behavior in the intermediate regime.
A new quantifier, the ratio of the amount of solute in dominantly advective versus dominantly diffusive pore channels, plays the role of `order parameter' of this phase transition.
Taylor dispersion, often invoked to explain the supra-linear behavior of longitudinal dispersion in this regime, was found not to be of primary importance.
The novel treatment of the intermediate regime paves the way for a more accurate description of dispersion as a function of flow velocity, spanning the whole range of P\'eclet numbers relevant to practical applications, such as ground water remediation.
\end{abstract}


\maketitle


\clearpage
\section{Introduction}

Transport of dissolved solutes in pore fluids in homogeneous porous media results from the synergy between advection and diffusion~\cite{Bear:1972,Marsily:1986}. 
Despite its importance in many applications and fields of research (e.g., groundwater remediation), this process is not yet understood in full detail. Flow through a porous medium causes an increased effective diffusion of the solute, termed dispersion, due to variations in flow velocity within and between the individual pore channels and due to the tortuous pathways the fluid follows through the pores. 
The corresponding dispersion coefficients are commonly applied in models based on the advection-diffusion  equation (ADE) to describe spreading of solute in porous media, for example pollutant plumes in ground water, although one should be aware that the ADE is only valid under rather restrictive conditions  \cite{Berkowitz:2002}.
Although it is generally agreed that dispersion increases with flow velocity, there is no agreement on the exact relationship between the dispersion coefficient and controlling parameters, such as flow velocity, pore geometry, fluid viscosity, etc.~\cite{Delgado:2006}. 

To study solute transport in homogeneous porous media, a sample of fluid-filled porous material is subjected to an external pressure head in a specific direction (taken to be the $x$-direction). This results in a fluid flow velocity $v_0$ through the medium, usually expressed in terms of the P\'eclet number, Pe $=v_0G/D$, where $G$ is a typical microscopic length scale (grain size), and $D$ a (`molecular') diffusion coefficient (of the solute in the fluid).
Then, the dispersion of an initially concentrated distribution of solute is studied as it is advected through the medium, while simultaneously experiencing diffusion.

The observed longitudinal ($D_x$) and transverse ($D_y$) dispersion as a function of P\'eclet number (flow velocity)
is often described by a disjunct set of up to five dispersional regimes~\cite{Fried:1971,Marsily:1986,Delgado:2007,Wood:2007}, using a separate functional relationship (sometimes referred to as `correlation') between P\'eclet number and dispersion, $D_{x,y}({\rm Pe})$, in each regime.
Typically, the following regimes are discerned, although the boundaries between the regimes vary somewhat between authors: 
(i) The molecular diffusion regime (Pe $< 0.1-0.3$);
(ii) the transition regime ($0.1-0.3 <$ Pe $< 5$); 
(iii) the major regime (also known as the power law regime~\cite{Bijeljic:2007}) ($5 <$ Pe $< 250-4000$); 
(iv) the mechanical dispersion regime ($250-4000 <$ Pe);
(v) the high Pe number regime, sometimes called the inertial or turbulent regime.
Here, we are concerned with laminar flow only. 
We will jointly refer to regimes (ii) and (iii) as the `intermediate regime', i.e., intermediate between the diffusive and mechanical dispersion regimes.

Quite often, the dispersion in each regime is described by a power law, i.e., $D_{x,y}({\rm Pe}) \propto {\rm Pe}^{\alpha_{x,y}}$. Not surprisingly, for regime (i), $\alpha_{x,y}=0$, while for regime (iv), $\alpha_{x,y}=1$. However, in the intermediate regime, exponents differing from these limiting values have been reported to fit experimental  data, which has given rise to speculation about their origin~\cite{Bijeljic:2004,Wood:2007}. Reported values for $\alpha_{x}$ are $1.2-1.3$ (\cite{Bijeljic:2004} and references therein; \cite{Wood:2007}), while those for  $\alpha_{y}$ are typically around $0.5-0.7$ \cite{Carvalho:2000,Klenk:2002,Olsson:2007}. 

Recently, a simplified heuristic model was proposed to replace this disjunct description of dispersion by a single, unified expression~\cite{Milligen:2012c,Bons:2013}.
The model assumes that the advective and diffusive transport mechanisms compete in the pore channels.
Then, as the mean flow velocity (or pressure head) is increased, transport in ever more pore channels along the solute flow path through the medium will be advection-dominated.
By making a simple assumption regarding the growth of the ratio between advection and diffusion dominated channels as the flow is increased, an expression for the net dispersion was derived. The expression successfully describes experimental data for dispersion in homogeneous porous media over the full range of P\'eclet numbers in laminar flow (regimes (i) to (iv)). In particular, it reproduces the faster than linear growth of the longitudinal dispersion with P\'eclet number in the intermediate regime, corresponding to an apparent exponent $\alpha_x>1$. 
It was claimed that this behavior could be understood from the statistical behavior of tracers in the pore channels.

To clarify the origin of this purported statistical behavior, presumed quite generic for porous media, 
here we study a highly simplified model for porous media consisting of a network of (pore) channels~\cite{Sahimi:1983,Sahimi:1988}.
In order to obtain a clear vision of the impact of the statistical behavior mentioned above on dispersion, the model we chose for this study is as minimalistic as possible, removing any physical mechanisms, such as turbulence and Taylor dispersion (see Section \ref{Motivation} below), that might affect these statistical properties. 
Thus, along the network connections, transport is one-dimensional and strictly diffusive and/or advective.
The effective longitudinal and transverse dispersion coefficients are extracted from the final numerical solution, after evolving the system in time.
It will be shown that this model does indeed reproduce the dispersion regimes and produces exponents $\alpha_{x,y}$ very similar to those obtained in experiments on actual porous media.
In this way, the minimum ingredients giving rise to the observed dispersional behavior are identified.
Furthermore, we will extract statistical information regarding the microscopic transport process that will elucidate the origin of the observed behavior.

{Clearly, the model has only limited relevance as de detailed model of real systems.
However, we emphasize that this is not its purpose. Rather, the model is constructed
to discriminate sharply between qualitatively and quantitatively different physical mechanisms.
Discrimination is  achieved by the combination of several assumptions, namely:}
(a) Highly complex three-dimensional porous materials are modeled by a simple two-dimensional network of infinitely thin connections linking nodes.
(b) Fluid flow through the system is imposed and not influenced by the presence of the solute; in other words, the solute fraction is assumed to be infinitesimally small.
(c) The fluid flow itself is incompressible, which is a reasonable assumption even in a realistic porous system when the fluid chosen is water or similar.
(d) The solute is passively transported by the fluid and is not assumed to be subject to independent transport equations (i.e., the solute has no inertia and it is not reactive).

This simplified model is used to study the effect of network topology on dispersion. 
The philosophy of our approach is similar to that of~\cite{Bijeljic:2004,Bijeljic:2007}.
{
Here, however, we render the model minimalistically.  The objective is to expose the essential
ingredients for description of solute transport in porous media.  An important aspect of the
model is the use of continuum transport equations for the solute. This effectively is the use
of an infinite number of tracers, which leads to high accuracy results (not easy to obtain
using tracers\cite{Rhodes:2006}).  Another important aspect is that the numerical model is specifically
designed to handle the wide spread of flow velocities in individual channels, typical of general porous media.
}

\section{The motivation of the simplified model approach}\label{Motivation}

{To motivate the model,} 
we briefly review the main mechanisms thought to cause dispersion in porous media~\cite{Fried:1971,Sahimi:1983,Marsily:1986}.

\subsection{Mechanical dispersion}

Fluid flows through a network of pore channels; we will only consider laminar flow.
Then, fluid flow is determined completely by the applied pressure head and the boundary conditions, e.g., no-slip boundary conditions at the channel walls. An important observation is that the whole problem of obtaining the fluid flow in the complex geometry and with given boundary conditions is {\it linear} in the applied pressure head: raising the head by a factor $f$ will lead to an increase of fluid velocity by the same factor $f$ everywhere.

Tracers are released into this fluid flow in a small region in space and time, and the tracer cloud is advected passively by the flow. 
We assume that the tracers are infinitesimal and massless (no inertia) so that they do not interact with each other, do not affect the flow, and follow the flow lines in the absence of diffusion. Further on, we will also consider the effect of (molecular) diffusion, but first we discuss pure flow effects. Tracers (and the fluid itself) cannot leave the network (particle conservation), except at the edge of the model network.

The tracer cloud, traveling through the network, will spread out due to the complex distribution of connections between nodes (leading to a complex flow pattern).
Note that this statement implicitly assumes that the flow through the network is such that the tracer cloud will actually spread out, i.e., that tracers may follow alternative paths leading to different net travelled distances from the point of injection -- this excludes, e.g., homogeneous flows ($v=$ constant over all space) from the analysis. 

After some time $t$, sufficiently large for initial transient effects to die out, but not so large that tracers are lost from the system, the size of the tracer cloud can be estimated by its spread 
\begin{equation}
\langle d_x^2\rangle = \langle (x - \langle x \rangle) ^2 \rangle.
\end{equation}
Here, $x$ indicates the set of $x$-coordinate values of the tracers, and the angular brackets imply a mean over all tracers.
Similar expressions hold for the spread in the other coordinate directions.
The corresponding effective dispersion coefficient can be estimated from 
\begin{equation}
D_v^x = \frac{\langle d_x^2\rangle}{t}
\end{equation}
in the $x$ direction, and similar for the other coordinate directions, assuming the initial size of the tracer cloud is infinitesimally small. 
Note that we call this dispersion coefficient {\it effective}, as the tracer distribution may deviate from a Gaussian shape in specific pore geometries. 
Deviations from Gaussianity may indicate that the ADE is an unsatisfactory model for global dispersional behavior \cite{Berkowitz:2002}. 
{In spite of this, the foregoing effective dispersion
coefficient can always be evaluated in finite-size systems at finite times. 

As noted, an increase of the pressure head by a factor $f$ increases the flow velocity everywhere by that same
factor, $v'=fv$.}
As we have limited ourselves to laminar flow and exclude turbulence and inertia, tracers will traverse exactly the same paths as they travel through the same porous medium from the same injection point in the new velocity field, but they will travel at a higher speed and hence complete their trajectories in less time, $t'=t/f$.
Thus, we can calculate the dispersion at a different pressure head as follows:
\begin{equation}
D_{v'}^x = \frac{\langle d_x^2\rangle}{t/f} = f D_v^x
\end{equation}
Using $f=v'/v$, one obtains $D_{v'}^x/v' = D_v^x/v$ for all $\{v', v\}$, i.e., $D_v^x = \beta^x v$.
In other words, this dispersion (commonly called `mechanical' dispersion) is proportional to the imposed flow velocity $v$ or the applied pressure head.
Note that this fundamental property of mechanical dispersion does not depend on the complexity of the flow (although excessively simple flows are excluded from this argument -- see above); all that is required is that the flow velocity everywhere depends linearly on the applied pressure head. 

\subsection{Diffusion}

In addition to pure passive advection by the fluid flow, the tracers will also experience a molecular diffusion, the rate of which is determined by the diffusion coefficient, $D$, which we will assume to be constant.
Diffusion is essentially distinct from mechanical dispersion in the sense that it is isotropic and non-deterministic.

\subsection{Taylor dispersion}\label{Taylor}

An additional effect contributing to {total} dispersion is Taylor dispersion~\cite{Taylor:1953}. 
In a flow field with shear (such that neighboring flow lines have different velocities), diffusion may lead to a homogenization of the velocity field, leading to an increased dispersion. 
This effect occurs only when the mean flow velocity is small with respect to the diffusion coefficient, i.e., when the P\'eclet number is sufficiently small.
In his original paper, Taylor derived an explicit expression for this condition, applicable to laminar flow through a straight pipe; namely
Pe $=Lv/D \ll 3\cdot 8^2 L^2/a^2$ (here, $L$ and $a$ are the length and radius of the pipe, respectively).
This condition { typically is} met over most of the intermediate dispersion regime of interest here.
Provided this condition is satisfied, the resulting dispersion is proportional to the {\it square} of the velocity (or P\'eclet number).

Taylor dispersion {often is} invoked to  explain {how} the longitudinal dispersion exponent $\alpha_x$ may exceed 1 in the intermediate regime in porous media~\cite{Bruderer:2001,Bijeljic:2004,Wood:2007,Aggelopoulos:2007,Charette:2007}.
{
The main motivation for this claim appears to be
that Taylor dispersion is one of the few available physical mechanisms
that provides faster than linear growth (of the dispersion) with Pe number.
To clarify whether Taylor dispersion is necessary for the description of dispersion
in homogeneous porous media, the simplified model we present below
provides a sharp discrimination among causes.  It is constructed such that
there are no neighboring flow lines that might lead Taylor dispersion
in the presence of diffusion (enhanced effective diffusion due to flow velocity
differences within a channel or network node).  Therefore, if the model
reproduces the dispersional behavior observed experimentally, Taylor dispersion may
not be construed as a necessary ingredient of dispersion models for homogeneous porous media.}

{
While the underlying model itself is not new, the simplification is, and the resulting
analysis and  interpretation are also. Specifically we are able to: (a) show
the evolution of the dispersion coefficients with P\'eclet number over a very broad
range and explain the origin of this behavior; (b) clarify the contribution
of Taylor dispersion; and (c) report the observation of a phase transition. In
the future, the proposed analysis and interpretation described here may be
applied to more complex and realistic models.}

\section{Numerical solution of advective + diffusive transport in a network}

{The model consists of a network of straight, one-dimensional  flow channels which
connect nodes.  Each point in the network  is characterized by a single velocity, unaffected by neighboring velocities. 
We model global transport across the network due solely to advection and diffusion. }
Along each connection, the diffusion coefficient $D>0$ and the advective velocity $v$ are known and constant.
Transport along each such a connection is therefore described by the one-dimensional transport equation
\begin{equation}\label{transport}
\frac{dp}{dt} = D \frac{d^2p}{dz^2} - v \frac{dp}{dz} 
\end{equation}
Here, $z$ is a local coordinate along the connection, while $p$ is the probability (or concentration).
The flux is given by
\begin{equation}\label{flux}
F = -D \frac{dp}{dz} + v p
\end{equation}
so that Eq.~(\ref{transport}) can also be written $dp/dt = -dF/dz$.

The system is discretized such that $p$ is only known at the nodes of the network, $p_i,i=1,\dots,N$.
Starting from an initial distribution $p_i = p_i^0$, we wish to evolve $p_i$ in time across the whole network, implying that we need to find $dp_i/dt$ at the nodes. To avoid probability leakage between the discrete nodes, the flux must be conserved in-between nodes, i.e., along each connection. Hence $dF/dz=0$ along the {channels} (though not at the nodes). 
Thus, from Eq.~(\ref{flux}) one obtains the shape of the solution along the {channel}:
\begin{equation}\label{solution}
p(z) = \left\{\begin{array}{cc}A\frac{D}{v}\left (e^{vz/D} -1\right ) + B & (v \ne 0) \\
Az + B & (v=0)\end{array}\right.
\end{equation}
with two integration constants $A$ and $B$.

A two-dimensional network consists of $N$ nodes at positions $\{x_i,y_i\}$, with $i=1,\dots,N$. 
A higher dimensional network simply means using a larger number of coordinates; 
all the other considerations and calculations remain essentially the same.
An $N \times N$ connection matrix $c_{ij}$ specifies to which other nodes each node $i$ is connected: $c_{ij}=1$ if node $j$ is connected to node $i$, and $c_{ij}=0$ otherwise ($c$ is a symmetric matrix without trace, $c_{ii}=0$).
In view of the fact that nodes are only connected to near neighbors, $c_{ij}$ is a sparse matrix. 
{Sparse matrix techniques therefore are appropriate.}

Now consider the connection between interior node $i$ and node $j$ (assuming $c_{ij}=1$).
Without loss of generality, we can define the local connection coordinate $z$ so that $z=0$ at node $i$. 
The length of the connection, $z_{ij}$, is (in two dimensions):
\begin{equation}
z_{ij}^2 = (x_j-x_i)^2+(y_j-y_i)^2
\end{equation}

Assuming the values of the solution $p$ at the two end nodes $i$ and $j$ are known, one may deduce $A_{ij}$ (the value of $A$ associated with the connection) from Eq.~(\ref{solution}):
\begin{equation}\label{A}
A_{ij} = \frac{(p_j-p_i)}{z_{ij}} \zeta \left ( \frac{v_{ij}z_{ij}}{D_{ij}}\right )
\end{equation}
and $B_{ij}=p_i$, where $D_{ij}$ is the diffusivity of the channel and $v_{ij}$ the velocity, and
\begin{equation}\label{zeta}
\zeta(z) = \left\{\begin{array}{cc}\frac{z}{\left (e^{z} -1\right )} & (z \ne 0) \\
1 & (z=0)\end{array}\right.
\end{equation}
The function $\zeta(z)$ is continuous for all $z$.

From Eq.~(\ref{flux}), the flux at node $i$ associated with the connection to node $j$ is found:
\begin{equation}\label{F}
F_{ij}= v_{ij}p_i -D_{ij} \left . \frac{dp(z)}{dz}\right |_{z=0}= v_{ij}p_i  - D_{ij} A_{ij}
\end{equation}

Combining Eqs.~(\ref{A}) and (\ref{F}):
\begin{equation}\label{F1}
F_{ij}=  v_{ij}p_i -\frac{(p_j-p_i)D_{ij}}{z_{ij}} \zeta \left ( \frac{v_{ij}z_{ij}}{D_{ij}} \right ) 
\end{equation}

The time rate of change of the solution at  node $i$ can be obtained from the continuity equation,
$\dot p = -\nabla \cdot \vec F$, i.e., $\int{\dot p dV}= -\int {F\cdot dS}$, so:
\begin{equation}
V_i\frac{dp_i}{dt} = -\sum_{j}{c_{ij}S_{ij}F_{ij}}
\end{equation}
Here, $V_i$ is the `volume' associated with node $i$, and $S_{ij}$ the `cross section' of channel $ij$.
Defining  $V_i = z_{ij} S_{ij}$, one obtains
\begin{equation}\label{evolution}
\frac{dp_i}{dt} = -\sum_{j}{c_{ij}\frac{F_{ij}}{z_{ij}}}
\end{equation}

Contrary to standard finite difference techniques, the present approach allows handling situations with greatly varying values of the local velocity $v_{ij}$ without resorting to very small time steps and/or node distances to avoid loss of accuracy (as discussed in~\cite{Heaton:2012}, Section 2.5), due to the fact that no probability leakage occurs along the connections, regardless of the velocity (cf.~next section).
Thus, the finite difference update of $p$ (i.e., $p_i^{\rm new} = p_i + \Delta t \cdot dp_i/dt$) is inherently stable.
This is particularly important for the present study, as the global transport behavior is explored for a rather wide range of advective velocities.

By default, edge nodes satisfy Neumann boundary conditions (zero outgoing flux).
However, we contemplate indicating specific edge nodes $i$ at which Dirichlet boundary conditions should hold, by means of a vector $e$ such that $e_i=1$ for Dirichlet condition nodes and $e_i=0$ otherwise. A second vector $p^e_i$ will then allow specifying the value of the solution at the nodes $i$ with $e_i=1$.
At such nodes, one has $dp_i/dt=0$ and:
\begin{equation}\label{boundary}
p_i = p_i^e \qquad (e_i=1)
\end{equation}

In addition, we will specify an initial condition $p_i(t=0) = p_i^0$. 
The simplest possible choice is a delta function, i.e., to set $p_i^0=0$ everywhere except at a specific node $i_0$ located near the centre of the grid, where we set $p_{i_0}^0=1$. This implies the existence of an initial transient, and time evolution should be followed long enough for this initial transient to die out. 

Using Eq.~(\ref{F1}), Eq.~\ref{evolution} can be written in matrix form as:
\begin{equation}\label{dpdt}
\frac{dp_i}{dt} = \sum_{j}{H_{ij} \cdot p_j}
\end{equation}
where 
\begin{eqnarray}\label{H}
H_{ij} = \left\{\begin{array}{cc}
c_{ij}\frac{D_{ij}}{z_{ij}^2}\zeta \left ( \frac{v_{ij}z_{ij}}{D_{ij}} \right ) & (e_i=0,i\ne j)  \\
-\sum_k{c_{ik}\left ( \frac{v_{ik}}{z_{ik}}+\frac{D_{ik}}{z_{ik}^2} \zeta \left ( \frac{v_{ik}z_{ik}}{D_{ik}} \right ) \right )} & (e_i=0,i= j) \\
0 & (e_i=1) \\
\end{array}\right.
\end{eqnarray}
The sparse matrix $H$ is constant over the time integration process, so that time integration is very efficient.
Time integration is carried out using a standard Runge-Kutta algorithm.

\subsection{Global probability conservation}

Eq.~(\ref{H}) has an important property: namely, total probability is conserved (assuming there are no Dirichlet nodes, i.e., $e_i=0$ for all $i$).
The total probability (or total `mass') $P$ is defined as
\begin{equation}
P = \sum_i{p_i}.
\end{equation}
From Eq.~(\ref{dpdt}) follows:
\begin{equation}\label{dlargepdt}
\frac{dP}{dt} = \sum_i{\frac{dp_i}{dt}} = \sum_{ij}{H_{ij} \cdot p_j}
\end{equation}
Expanding Eq.~(\ref{dlargepdt}) using Eq.~(\ref{H}):
\begin{eqnarray}
\frac{dP}{dt} =& \sum_{i\ne j}{\frac{c_{ij}D_{ij}p_j}{z_{ij}^2} \zeta \left ( \frac{v_{ij}z_{ij}}{D_{ij}} \right )} \nonumber \\
&-\sum_{ik}{c_{ik}\left ( \frac{v_{ik}}{z_{ik}}p_i+\frac{D_{ik}p_i}{z_{ik}^2}\zeta \left ( \frac{v_{ik}z_{ik}}{D_{ik}} \right )\right )}
\end{eqnarray}
In view of the fact that the local $z$ coordinate is always increasing from $i$ to $j$, $v_{ij}=-v_{ji}$ (antisymmetry).
On the other hand, $c_{ij}=c_{ji}$, $D_{ij}=D_{ji}$, and $z_{ij}=z_{ji}$ are symmetric.
It is convenient to split the double sums in two halves ($i>j$ and $i<j$, with opposite signs for $v$) to obtain cancellations.
Using the cited symmetry properties:
\begin{eqnarray}
\frac{dP}{dt} &=& \sum_{i>j}{\frac{c_{ij}D_{ij}p_j}{z_{ij}^2}\left [\zeta \left ( \frac{v_{ij}z_{ij}}{D_{ij}} \right )+\zeta \left ( \frac{-v_{ij}z_{ij}}{D_{ij}} \right ) \right ]}  \nonumber \\
&&-\sum_{k>i}{c_{ki}\frac{v_{ki}-v_{ki}}{z_{ik}} p_i} \nonumber \\
&&-\sum_{k>i}{\frac{c_{ki}D_{ki}p_i}{z_{ki}^2}\left [\zeta \left ( \frac{v_{ki}z_{ki}}{D_{ki}} \right )+\zeta \left ( \frac{-v_{ki}z_{ki}}{D_{ki}} \right ) \right ]}\nonumber \\ 
&=& 0
\end{eqnarray}
Q.E.D. This global property is beneficial for the global stability of the solution method.

\subsection{Velocity model}

In principle, the velocity $v_{ij}$ can be specified independently for each connection.
However, the systems we pretend to model are characterized by a constant mean global flow velocity in a specific direction (chosen to be the $x$-direction) caused by a global pressure drop across the system. The fluid is chosen to be incompressible (like water).
Incompressible flow is such that $\nabla \cdot \vec v = 0$, or $\vec v = -\nabla \phi$, where $\phi$ is a potential (pressure) field satisfying the Laplace equation, $\nabla^2 \phi = 0$.
Assume $\phi_i$ is known at all nodes. Then 
\begin{equation}\label{v}
v_{ij} = -(\phi_j-\phi_i)/z_{ij}.
\end{equation}
Assuming all connections have the same cross section, the condition $\nabla \cdot \vec v = 0$ translates into
\begin{equation}\label{incompressible}
\sum_{j}{c_{ij}v_{ij}} = 0
\end{equation}
for all nodes $i$.
Again, this can be written in matrix form, namely
\begin{equation}\label{G}
\sum_{j}{G_{ij}\phi_{j}} = \phi^e_i
\end{equation}
where
$\phi^e_i$ has a given value on the set of nodes $i \in \{E_L,E_R\}$ corresponding to the left and right edges of the grid (the potential values at those positions), and takes value zero for all other nodes, while
\begin{eqnarray}
G_{ij} = \left\{\begin{array}{cc}
-\frac{c_{ij}}{z_{ij}} & (i\ne j,i\notin \{E_L,E_R\}) \\
\sum_k{\frac{c_{ik}}{z_{ik}}} & (i= j,i\notin \{E_L,E_R\}) \\
\delta_{ij} & (i \in \{E_L,E_R\}) \\
\end{array}\right.\nonumber
\end{eqnarray}
where $\delta_{ij}$ is the Kronecker delta ($\delta_{ij}=1$ when $i=j$ and 0 otherwise).
The potential distribution $\phi_i$ is immediately obtained by solving Eq.~(\ref{G}) using linear least square techniques, after which $v_{ij}$ is found from Eq.~(\ref{v}).

If the potential of the right edge of the grid is kept at $\phi_i^e=0$, then the left edge will be at $-L_x v_0$, where $L_x$ is the total size of the grid in the $x$ direction, such that the mean flow is in the $x$-direction with velocity $v_0$.

\subsection{Effective dispersion coefficients}

In order to display the solution $p_i$, $i=1,\dots,N$ (at any given time $t$) and determine the effective dispersion coefficients, it is interpolated onto a regular and sufficiently fine $x-y$ grid, resulting in an interpolated distribution $p_{\rm int}(x,y)$. 
The effective diffusion coefficients (at the integration endpoint $t=T$) are computed from:
\begin{eqnarray}\label{diffusion}
D_x &=& \left \langle \left (x-\langle x \rangle \right )^2 \right \rangle  \mathbin{/} T, \nonumber \\
D_y &=&  \left \langle \left (y-\langle y \rangle \right )^2 \right \rangle  \mathbin{/} T,
\end{eqnarray}
where $\langle f \rangle$ is defined as the weighted mean of the quantity $f$ over $p_{\rm int}$:
\begin{equation}
\langle f \rangle = \frac{\int \!\! \int{f p_{\rm int}(x,y)~dxdy}}{\int \!\! \int{p_{\rm int}(x,y)~dxdy}}
\end{equation}
Of course, the diffusion coefficients, Eq.~(\ref{diffusion}), are {\it effective} as the final distribution is not necessarily Gaussian; 
the main point being that the diffusion coefficients in laboratory experiments are usually determined in a similar or equivalent manner.

\subsection{Transport quantifiers}

To understand global transport across the network, we define some auxiliary quantities.
Based on previous work~\cite{Milligen:2012c} we expect $t_v/t_0$, the average time spent by a tracer in advection-dominated channels in relation to the time spent in diffusion-dominated channels, to be of prime importance. 
To estimate this number, we must first define a quantifier that tells us which mechanism is dominant in a given channel. 
This quantifier is the `local P\'eclet number': 
\begin{equation}\label{peloc}
{\rm Pe}^{\rm loc}_{ij} = \frac{|v_{ij}|z_{ij}}{D_{ij}}. 
\end{equation}
When ${\rm Pe}^{\rm loc}_{ij}\le1$, diffusion is said to dominate in channel $ij$, and otherwise advection dominates.
At any time $t$ large enough such that the initial transient state has decayed, the density of the solute ($\overline {p_{ij}}$) in advective or diffusive channels ($n_v$ or $n_0$) is linearly proportional to the residence time of tracers in such channels 
($t_v$ or $t_0$).
The mean solute  density of channel $ij$ is (cf.~Eq.~(\ref{solution}) ff.):
\begin{equation}\label{meanp} 
\overline {p_{ij}} = p_i + (p_j-p_i)\frac{D_{ij}}{v_{ij}z_{ij}}\left ( 1- \zeta \left ( \frac{v_{ij}z_{ij}}{D_{ij}} \right )\right )
\end{equation}
Thus, we define the two numbers
\begin{eqnarray}
n_v = \sum_{{\rm Pe}^{\rm loc}_{ij}> 1}{\overline {p_{ij}}}  \nonumber \\
n_0 = \sum_{{\rm Pe}^{\rm loc}_{ij}\le 1}{\overline {p_{ij}}} 
\end{eqnarray}
The sums are conditional, i.e., they run over all connections $ij$ satisfying the condition. 

\subsection{Velocity correlation}

It is interesting to compute the correlation function between the solute motion and the flow velocity.
This correlation is expected to be low for small $v_0$ (where diffusion dominates), and high for high $v_0$ (where advection dominates).
In the case of a model with tracers, the calculation of such a correlation is done by correlating the instantaneous tracer velocity with the local flow velocity $v_{ij}$. However, the current model does not contain tracers. 
Even so, an effective solute velocity can be defined by
\begin{equation}
\vec v^s = -\frac{dp}{dt} \left ( \vec \nabla p \right ) ^{-1}
\end{equation}
i.e., along each connection $ij$:
\begin{equation}
v^s_{ij} = -\frac12 \left (\frac{dp_i}{dt}+\frac{dp_j}{dt}\right )  \frac{z_{ij}}{(p_j-p_i)}
\end{equation}
The correlation between $v^s_{ij}$ and the externally imposed flow $v_{ij}$ is found from
\begin{equation}\label{correlation}
C^{s,0} = \frac{\langle v^s_{ij}v_{ij}\rangle}{\left [ \langle (v^s_{ij})^2\rangle \langle (v_{ij})^2\rangle \right ]^{1/2}}
\end{equation}
where the angular brackets refer to a mean over all network connections, weighted by the mean solute density $\overline {p_{ij}} $, Eq.~(\ref{meanp}):
\begin{equation}
\langle g_{ij}\rangle  = \left . \sum_{ij}{c_{ij}\overline {p_{ij}}  g_{ij}} \middle / \sum_{ij}{c_{ij}\overline {p_{ij}} } \right .
\end{equation}

%
 
\clearpage
\section{Results}

In the following, we will set $D_{ij}=D=1$, physically corresponding to a homogenous fluid at a constant temperature.
The diameter of the network cells is typically $G=1$, although the definition of `cell diameter' is not unambiguous for all types of network.
With these choices, the P\'eclet number Pe $=G \cdot v_0/D$ is equal to $v_0$ itself, which is useful for the interpretation of the results below in relation to experimental data. 
A priori, these numbers have not yet been assigned a specific physical dimension.
However, if we use mm as the spatial unit, and s as the time unit, then $G$ is in mm and $D$ in mm$^2$/s, which actually makes these choices very close to the physical values used in many sand box or bead experiments.

Edge boundary conditions are the default ones (Neumann), although this is of little consequence as time integration is stopped when the distribution $p$ reaches the boundary.
We define a stopping criterion based on the ratio
\begin{equation}
R = \frac{\max_{\rm edge}(p)}{\max(p)}
\end{equation}
Initially, $R=0$ as the initial condition is $p_i=0$ for all $i$ except at one centrally located node.
As $p$ evolves in time, the distribution broadens and after some time the value of $p$ at the edge nodes begins to increase, such that $R$ increases.
We stop the time integration when $R>\theta$, where $\theta$ is a threshold.
Here, we set $\theta=0.001$.

The network size should be sufficiently large to allow any initial transient to decay.
Typically, network sizes of about 100 cells in each dimension should be more than sufficient, although this will depend somewhat on the specific grid topology.

Before discussing the numerical results, we {reiterate} that this model does not contain Taylor dispersion.
First, as transport through every network channel is one-dimensional, there are no neighboring flow lines with velocity differences that can be smoothed out by diffusion (see Section \ref{Taylor}).
Second, a hypothetical tracer particle crossing from one flow line to another at a network intersection node only experiences the local flow velocity at every time instant, so there is no mixing of flow velocities as would occur with the Taylor dispersion mechanism.
Third, along each network channel, the diffusion $D$ is constant, so there is also no `effective' Taylor dispersion (which one might introduce by defining $D \propto v^2$).

\clearpage
\subsection{Square grid}

\begin{figure}\centering
  \includegraphics[trim=0 0 0 0,clip=,width=12cm]{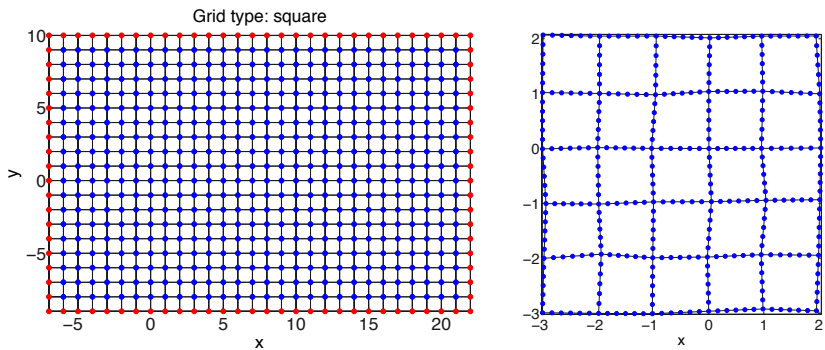}
\caption{\label{square_grid}Left: square grid $30 \times 20$. Edge nodes shown in red. Right: small section of this grid after `wiggling' and inserting additional sub-nodes (see text).}
\end{figure}

An example of a small square grid is shown in Fig.~\ref{square_grid}. Runs are done on a larger grid.
An issue affecting regular grids like these is the fact that many connections are exactly vertical.
Such channels, being perpendicular to the (horizontal) flow direction, are `stagnant', and at high velocities $v_0$ they cease to contribute to the transport, so that the effective transport becomes one-dimensional instead of two-dimensional. 
To avoid this topological breakdown from occurring, the main grid nodes are `wiggled':
each main grid point is moved randomly (in both $x$ and $y$) by an amount $f_w \cdot G \cdot \epsilon$, where $f_w=0.1$ is a `wiggle factor', $G$ is the mean distance between main grid points, and $\epsilon$ is a uniformly distributed random number in the interval $(-0.5,0.5)$.
After wiggling the main grid nodes (grid intersections), additional sub-nodes are placed equidistantly along each connection in order to resolve the spatial variation of the solution along the connections (cf. Fig.~\ref{square_grid}).

\begin{figure}\centering
  \includegraphics[trim=0 0 0 0,clip=,width=12cm]{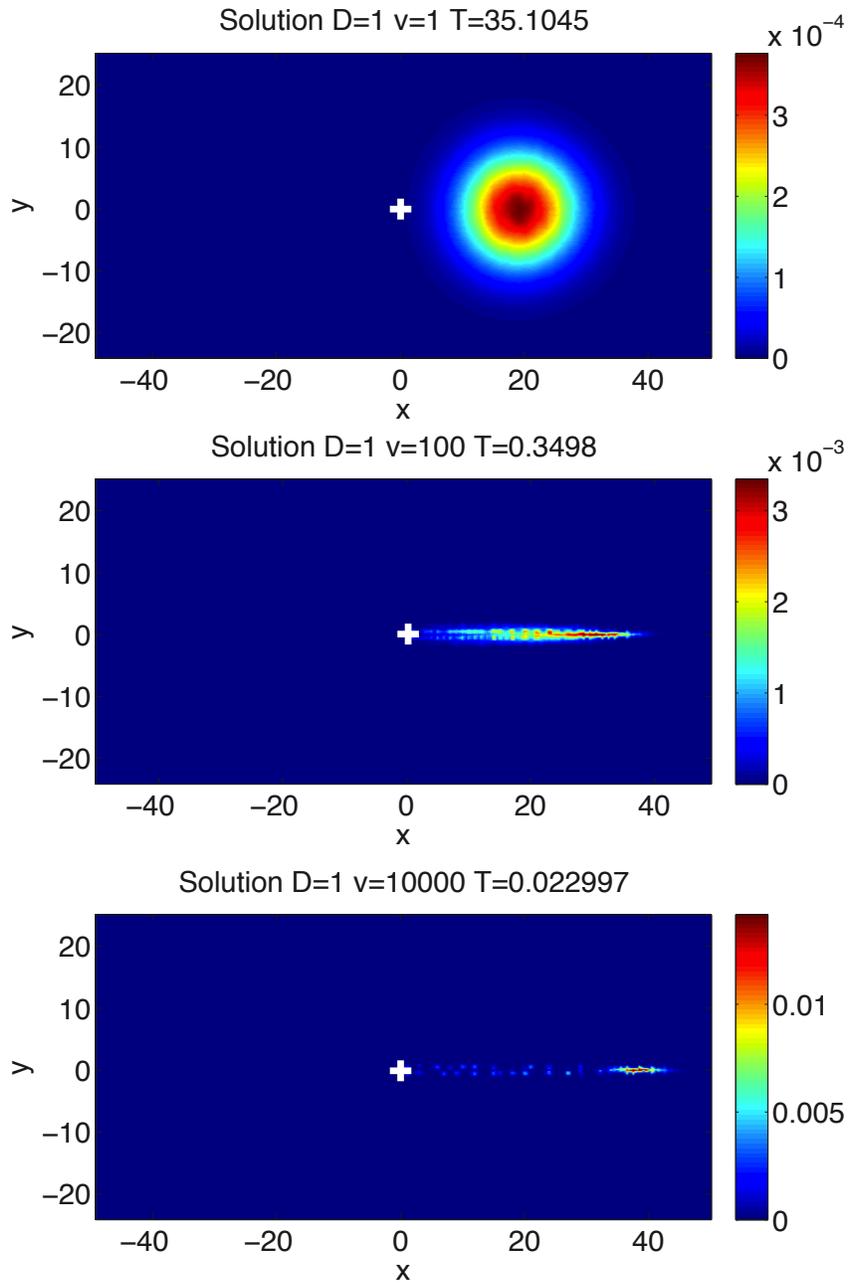}
\caption{\label{square_solution}Square grid $100 \times 50$, each connection between main nodes subdivided into 6 sub-segments.  
$p_{\rm int}(x,y)$ at $t=T$ for $v_0=1$ (top), $v_0=100$ (middle) and $v_0=10000$ (bottom).
The solute injection point (the location of the initial delta function) is marked by a cross.}
\end{figure}

Fig.~\ref{square_solution} shows an example of the solution $p_{\rm int}(x,y)$ at $t=T$ for $v_0=1,100,10000$.
The effective velocity of the bulk of the solute is less than the fluid flow velocity, $v_0$, as part of the forward movement is dissipated into the nearly perpendicular, diffusion-dominated channels.

\begin{figure}\centering
  \includegraphics[trim=0 0 0 0,clip=,width=12cm]{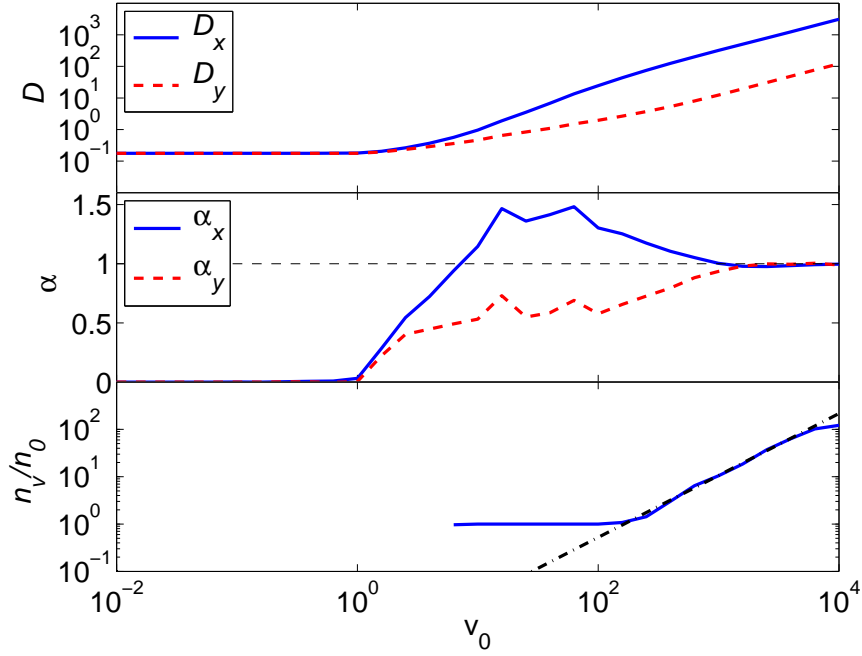}
\caption{\label{square}Square grid $100 \times 50$, each connection between main nodes subdivided into 6 sub-segments. Total number of nodes: 54250. Top: $D_x$ and $D_y$ versus $v_0$. Centre: $\alpha_x$ and $\alpha_y$. Bottom: $n_v/n_0$. The shown line is $n_v/n_0 \propto v_0^\gamma$ with $\gamma = 1.3$.
}
\end{figure}

Fig.~\ref{square} shows $D_x$ and $D_y$ versus $v_0$.
These curves resemble the experimental results~\cite{Delgado:2007} in various respects.
First, the limiting behavior is correct (in the diffusive regime, corresponding to small $v_0$, $D_{x,y}$ is constant; and in the mechanical dispersion regime, corresponding to large $v_0$, $D_{x,y} \propto v_0$). 
Second, the ratio $D_x/D_y$ approaches a number larger than 10 for high $v_0$ (namely, about 27).
Third, both $D_x$ and $D_y$ are significantly below 1 for small $v_0$, which is in accordance with experimental results and expectation: 
namely, the limitation that transport can only occur along the network connections means that effective diffusion along the network must be slower than diffusion through free space. Of particular interest is the fact that the longitudinal exponent $\alpha_x = d \ln D_x / d \ln v_0$ reaches a maximum value of 1.4 in the intermediate regime, in spite of the fact that this model does not contain Taylor dispersion. 
Fig.~\ref{square} also shows $n_v/n_0$: at $v_0 = 7$, this quantity changes from a value of 0 (no dominantly advective channels) to 1 (equipartition of dominantly advective / diffusive channels). This is also approximately the point where $\alpha_x$ rises above 1, suggesting that the high value of $\alpha_x$ is related to the increase of advection-dominated channels.

\clearpage
\subsection{Tile grid}

\begin{figure}\centering
  \includegraphics[trim=0 0 0 0,clip=,width=12cm]{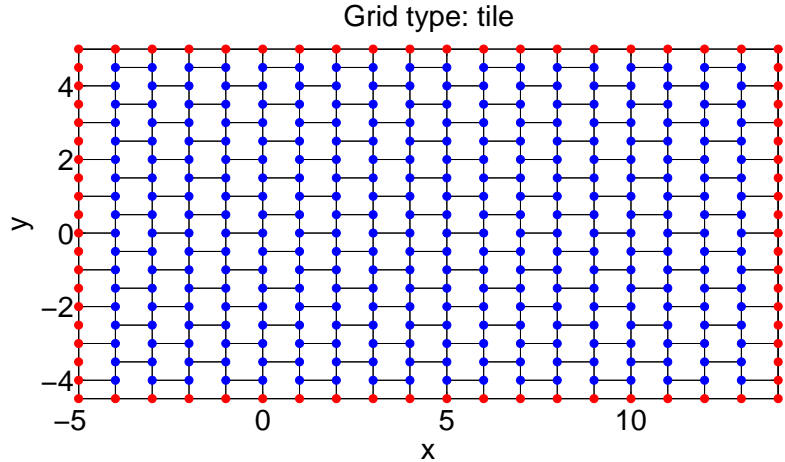}
\caption{\label{tile_grid}Tile grid $20 \times 20$. Edge nodes shown in red.}
\end{figure}

The `tile' grid is obtained from the square grid by omitting half of the horizontal cross connections and rescaling the vertical axis; as a consequence, it is strongly non-isotropic.
An example of a small tile grid is shown in Fig.~\ref{tile_grid}.

\begin{figure}\centering
  \includegraphics[trim=0 0 0 0,clip=,width=12cm]{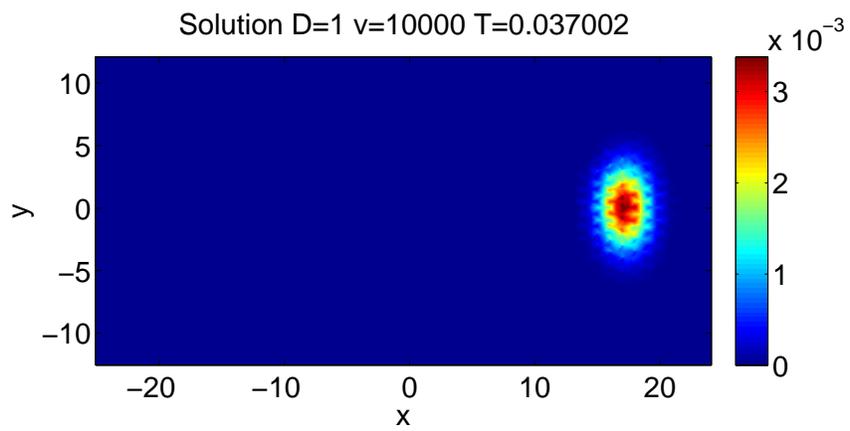}
\caption{\label{tile_solution}Tile grid $50 \times 50$, each connection between main nodes subdivided into 6 sub-segments. Total number of nodes: 20875.  
$p_{\rm int}(x,y)$ at $t=T$ for $v_0=10000$.}
\end{figure}

Runs are done on a larger grid.
As before, the main grid points were `wiggled' with factor $f_w = 0.1$.
Fig.~\ref{tile_solution} shows an example of the solution $p_{\rm int}(x,y)$ at $t=T$ for $v_0=10000$.
Comparing this result with the result for the square grid at high velocity, Fig.~\ref{square_solution}, it is seen that the solute `cloud' expands much more in the perpendicular ($y$) direction, as a consequence of the grid structure.

\begin{figure}\centering
  \includegraphics[trim=0 0 0 0,clip=,width=12cm]{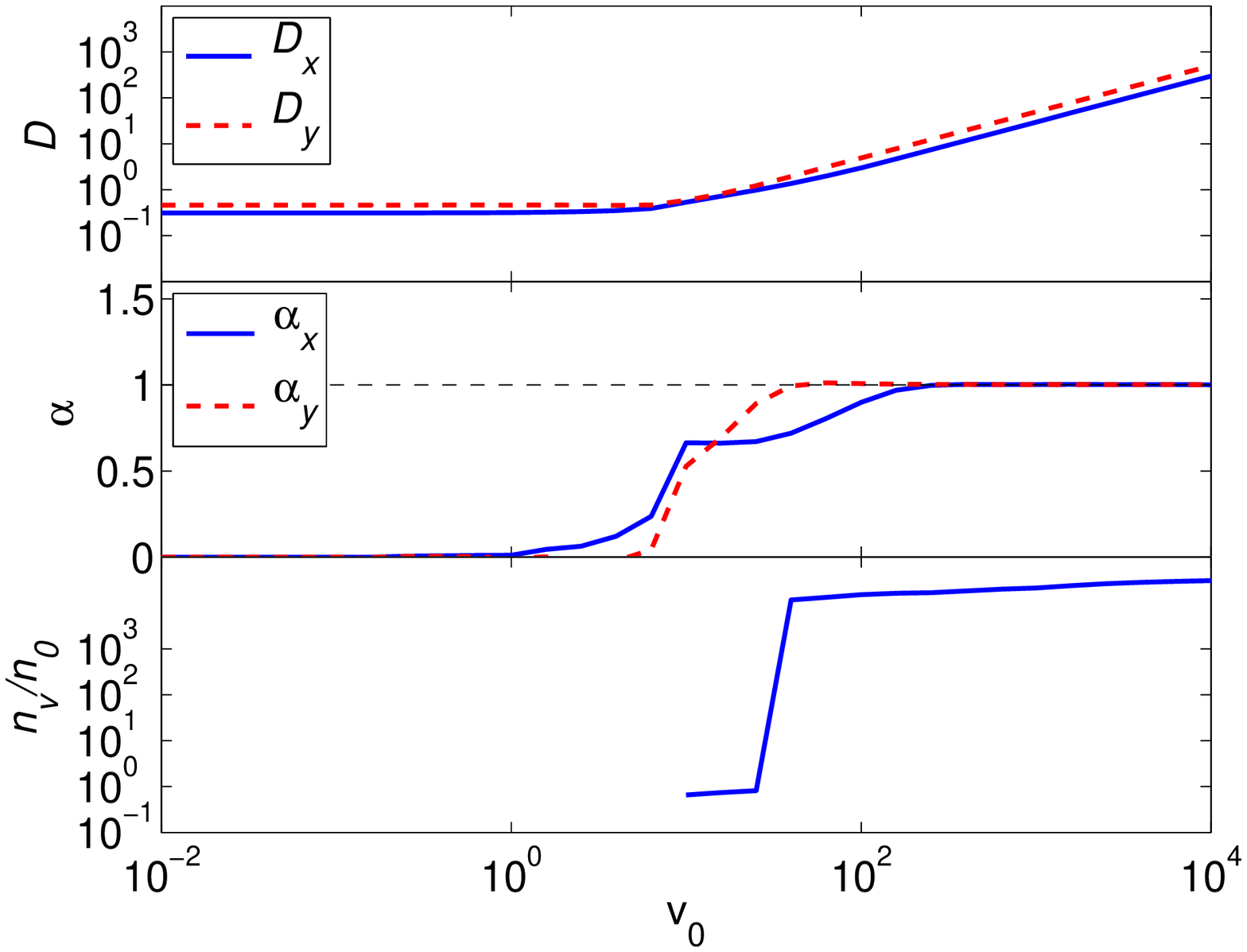}
\caption{\label{tile}Tile grid $50 \times 50$, each connection between main nodes subdivided into 6 sub-segments. Total number of nodes: 20875.  Top: $D_x$ and $D_y$ versus $v_0$. Centre: $\alpha_x$ and $\alpha_y$. Bottom: $n_v/n_0$.}
\end{figure}

Fig.~\ref{tile} shows $D_x$ and $D_y$ versus $v_0$.
This grid favors transport in the $y$-direction (the vertical channels contain no bends) and hinders horizontal transport (there are no through-going horizontal channels). Hence $D_y>D_x$, and even at large $v_0$, $D_x$ and $D_y$ remain very similar in size.
This example shows that the grid structure has a large influence on the final dispersion.

\clearpage
\subsection{Hexagonal grid}

\begin{figure}\centering
  \includegraphics[trim=0 0 0 0,clip=,width=12cm]{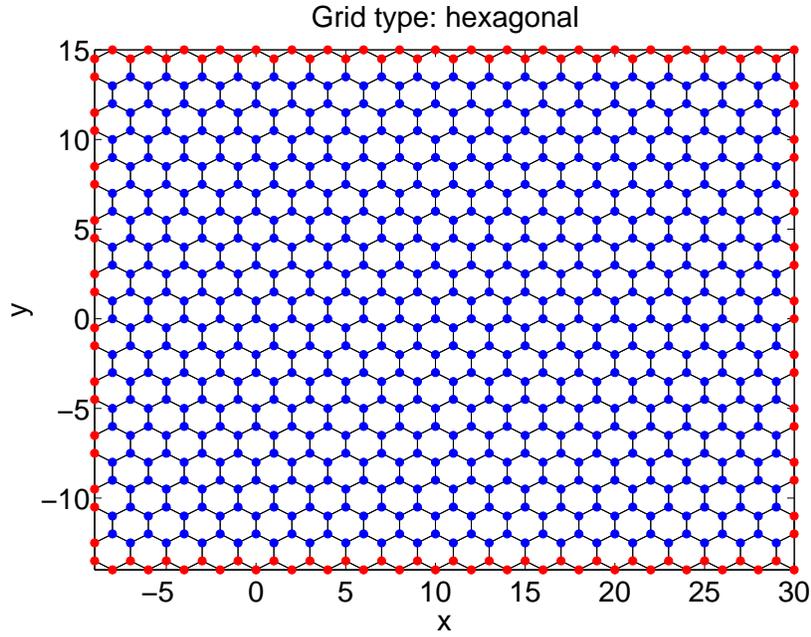}
\caption{\label{hexagonal_grid}Hexagonal grid $40 \times 20$. Edge nodes shown in red.}
\end{figure}

\begin{figure}\centering
  \includegraphics[trim=0 0 0 0,clip=,width=12cm]{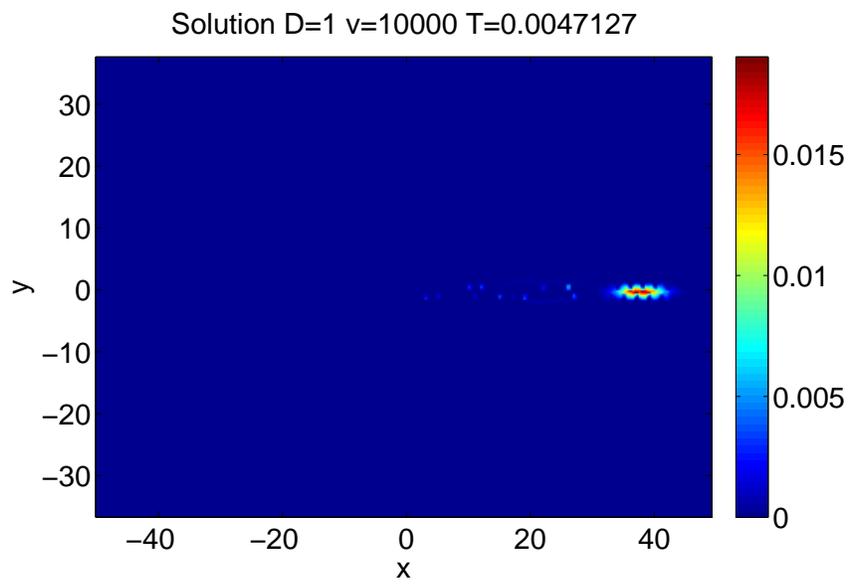}
\caption{\label{hexagonal_solution}Hexagonal $100 \times 50$ grid, each connection between main nodes subdivided into 6 sub-segments. Total number of nodes: 42000. 
$p_{\rm int}(x,y)$ at $t=T$ for $v_0=10000$.}
\end{figure}

An example of a small hexagonal grid is shown in Fig.~\ref{hexagonal_grid}.
Runs are done on a larger grid. As before, the main grid points were `wiggled' with factor $f_w = 0.1$.
A solution for high $v_0$ is shown in Fig.~\ref{hexagonal_solution}.

\begin{figure}\centering
  \includegraphics[trim=0 0 0 0,clip=,width=12cm]{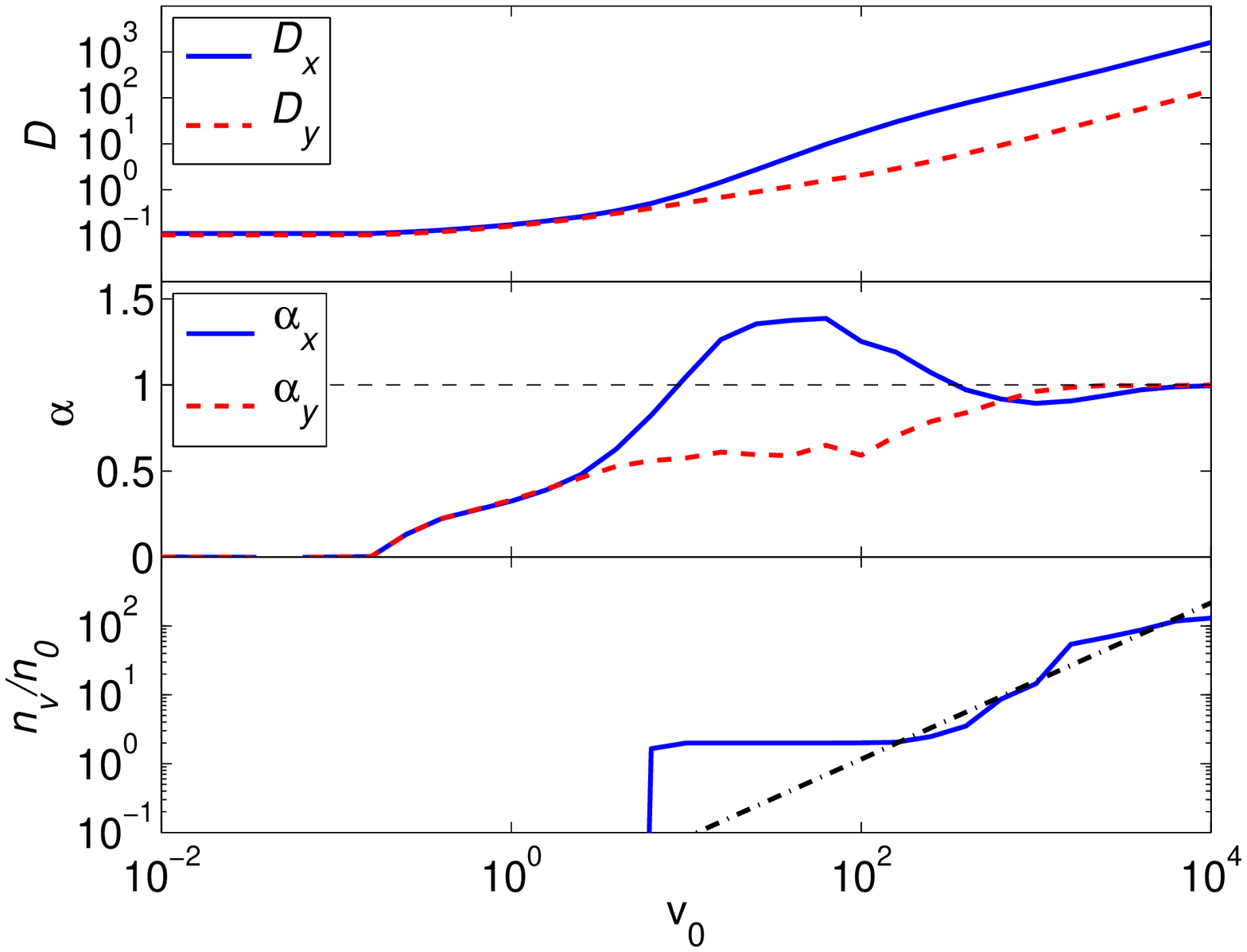}
\caption{\label{hexagonal}Hexagonal $100 \times 50$ grid, each connection between main nodes subdivided into 6 sub-segments. Total number of nodes: 42000. Top: $D_x$ and $D_y$ versus $v_0$. Centre: $\alpha_x$ and $\alpha_y$. Bottom: $n_v/n_0$. The shown line is $n_v/n_0 \propto v_0^\gamma$ with $\gamma = 1.1$.}
\end{figure}

Fig.~\ref{hexagonal} shows $D_x$ and $D_y$ versus $v_0$.
Note that $\alpha_x>1$ in the intermediate regime (maximum value: 1.4), and the relation between the $\alpha_{x,y}$ curves and the $n_v/n_0$ curve: at $v_0 = 7$, this quantity changes from a value of 0 (no dominantly advective channels) to 1 (equipartition of dominantly advective / diffusive channels). This is also approximately the point where $\alpha_x$ rises above 1, again suggesting that the increase of advection-dominated channels is related to the high value of $\alpha_x$.

\clearpage
\subsection{``Elle'' grid}

\begin{figure}\centering
  \includegraphics[trim=0 0 0 0,clip=,width=12cm]{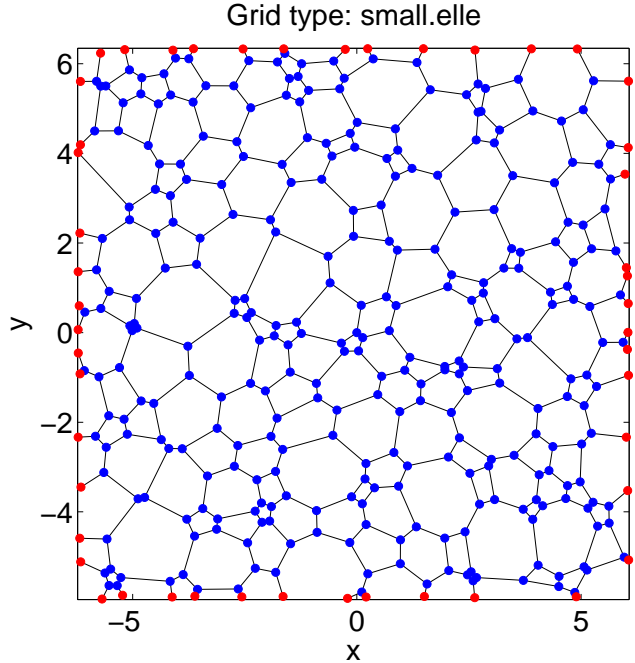}
\caption{\label{elle_grid}``Elle'' grid. Total number of nodes $N$: 344. Edge nodes shown in red. The initial node is at $(x,y)=(0,0)$.}
\end{figure}

In this section, we study a more generic, irregular network.
This network is a foam texture made with a routine for two-dimensional static grain growth obtained from the numerical simulation platform ``Elle''~\cite{Jessell:2001,Roessiger:2011}. 
This network is irregular on a small scale but homogeneous on a large scale.
An example of a small grid is shown in Fig.~\ref{elle_grid}.
The dimensions are such that the mean grain size is $d=1$.

\begin{figure}\centering
  \includegraphics[trim=0 0 0 0,clip=,width=12cm]{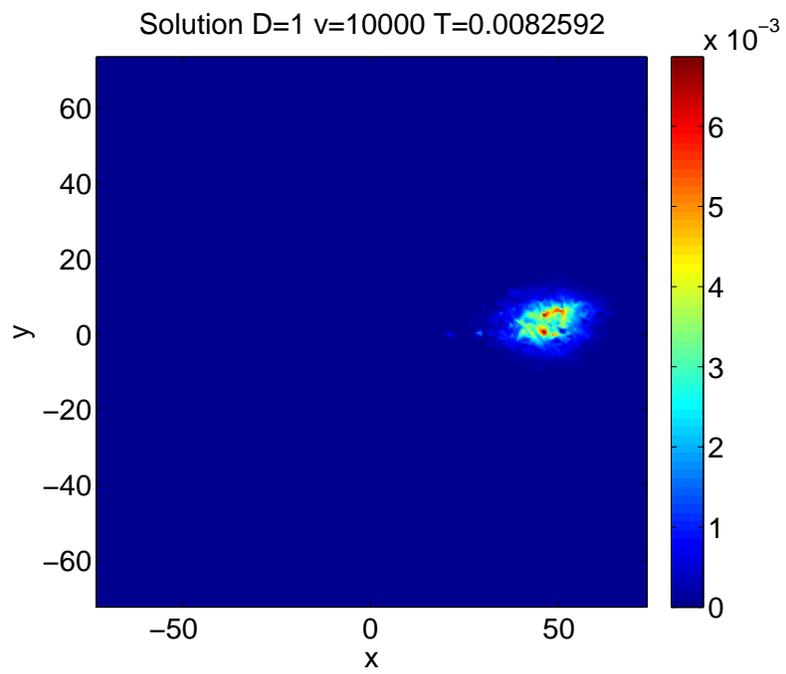}
\caption{\label{elle_solution}``Elle'' grid. $p_{\rm int}(x,y)$ at $t=T$ for $v_0=10^4$. Number of nodes: 31776.}
\end{figure}

\begin{figure}\centering
  \includegraphics[trim=0 0 0 0,clip=,width=12cm]{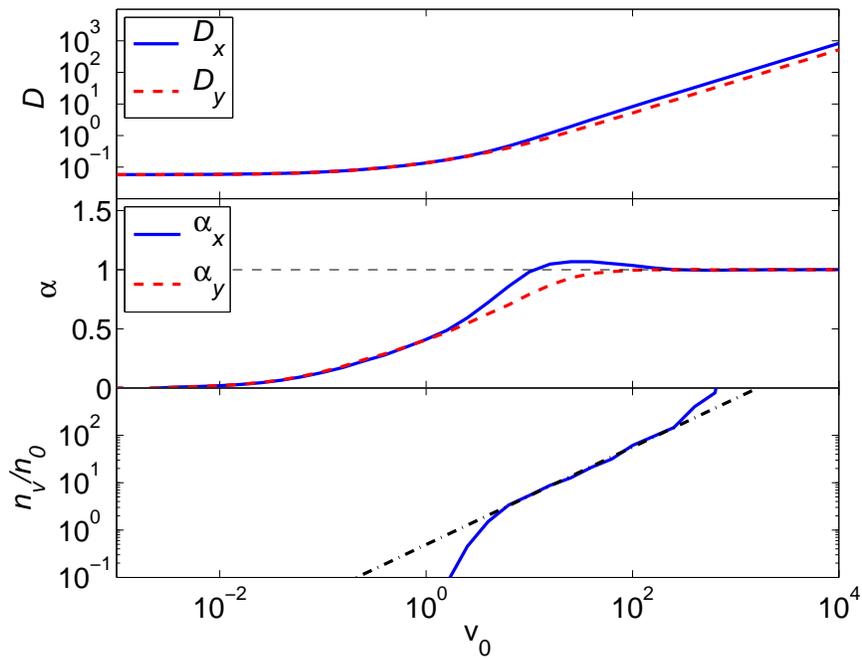}
\caption{\label{elle}``Elle'' grid. $D_x$ and $D_y$ versus $v_0$. Number of nodes: 31776.
Note that the local power-law exponent $\alpha_x>1$ in the intermediate regime. 
Also shown is $n_v/n_0$, along with a power law fit $n_v/n_0 = c v_0^\gamma, \gamma = 1.03 \pm 0.06$.}
\end{figure}

Runs are done on a larger network. 
Fig.~\ref{elle_solution} shows an example of the solution $p_{\rm int}(x,y)$ at $t=T$ for $v_0=10^4$.
The initial node is chosen more or less in the center of the grid.
Fig.~\ref{elle} shows $D_x$ and $D_y$ versus $v_0$.
Note that $\alpha_x>1$ in the intermediate regime (maximum value: 1.07).
As in the preceding examples, the $n_v/n_0$ curve reaches a value of 1 for values of $v_0$ slightly smaller than the values corresponding to the region where $\alpha_x>1$.
Here, however, the behavior of the $n_v/n_0$ curve is much smoother than with the regular grids.

\begin{figure}\centering
  \includegraphics[trim=0 0 0 0,clip=,width=12cm]{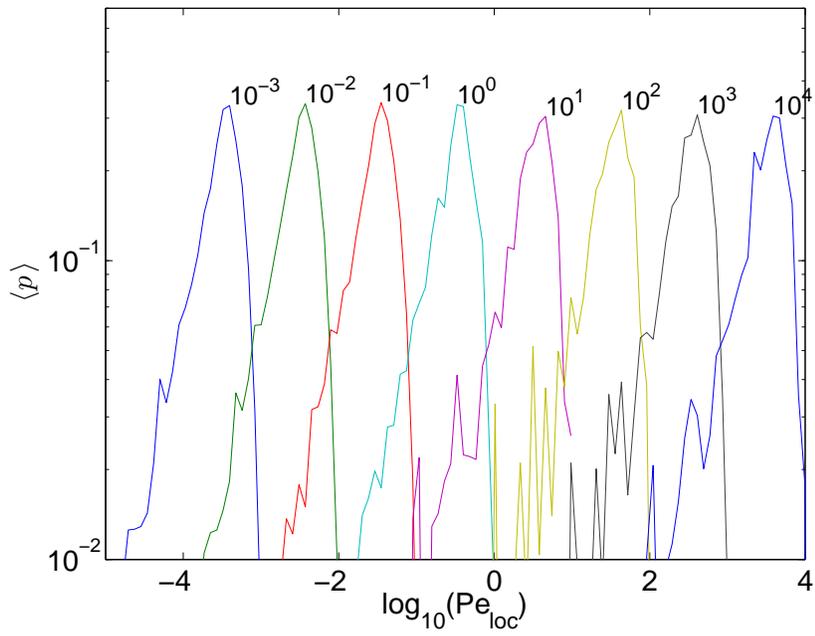}
\caption{\label{ellestats_2d}``Elle'' grid. Number of nodes: 31776.
Mean solute concentration in channels with given Pe$^{\rm loc}$ as a function of mean flow velocity $v_0$ (indicated by the curve labels).
The width of the distribution changes with $v_0$.
}
\end{figure}

To better understand the origin of this super-linear growth of the longitudinal dispersion coefficient, we have calculated the distribution of solute over channels with different Pe$^{\rm loc}$ values, Eq.~(\ref{peloc}), as a function of the fluid velocity $v_0$ (Fig.~\ref{ellestats_2d}).
Perhaps unsurprisingly, the distribution of solute shifts towards channels with higher Pe$^{\rm loc}$ as $v_0$ is raised, i.e., towards advection-dominated channels.
This shift is mainly due to the fact that the number Pe$^{\rm loc}$ for a given channel is linear in $v_0$ itself; obvious though this may be, it constitutes the essential ingredient of the explanation of $\alpha_x > 1$, discussed below.

\begin{figure}\centering
  \includegraphics[trim=0 0 0 0,clip=,width=12cm]{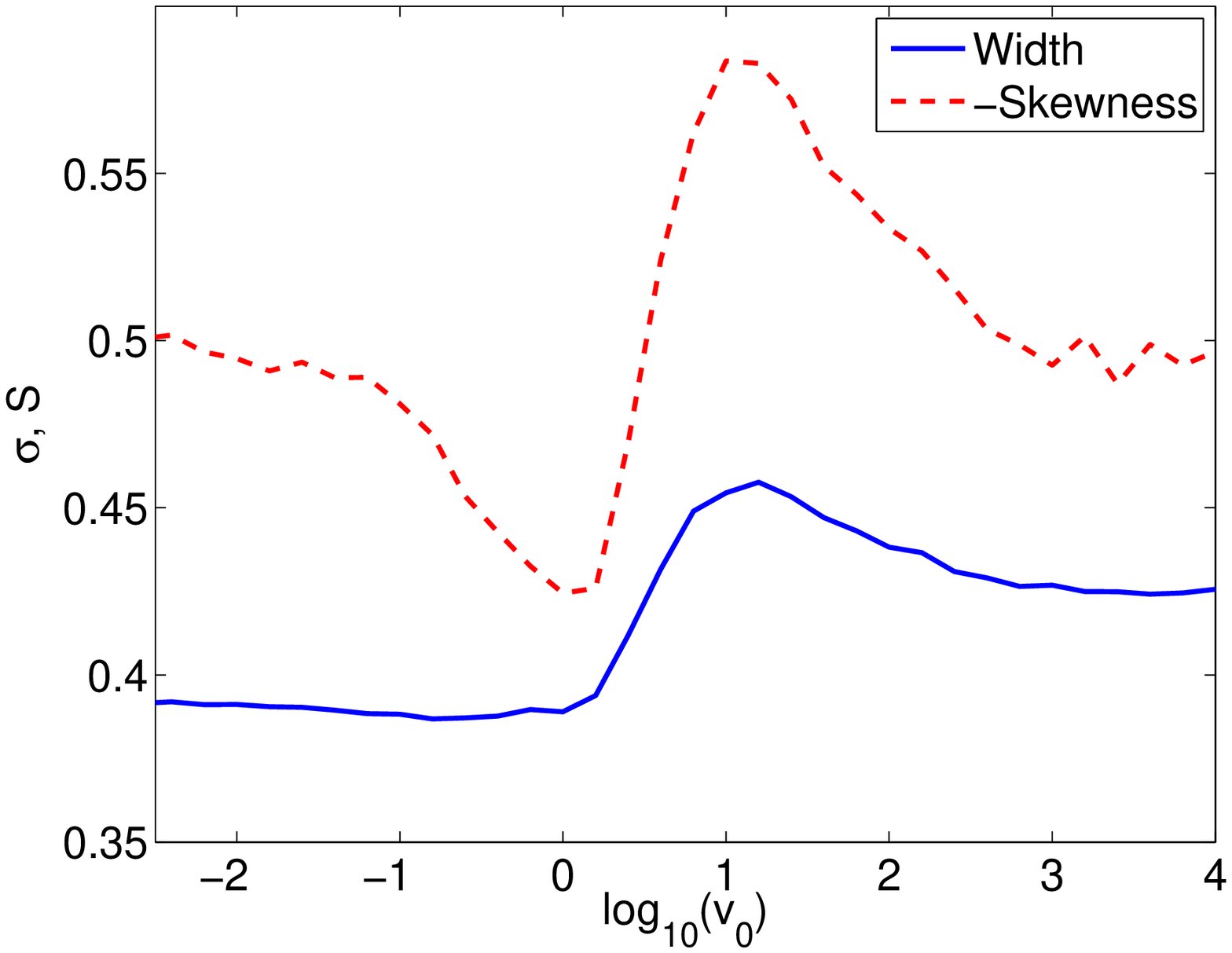}
\caption{\label{ellestats_width} ``Elle'' grid. Number of nodes: 31776.
The width $\sigma$ and skewness $S$ (shown with inverted sign for display purposes) of the distributions shown in Fig.~\ref{ellestats_2d}.
}
\end{figure}

Furthermore, a gradual change of the {\it shape} of the distribution is visible as $v_0$ is raised; however, this change is rather subtle. 
To quantify and visualize this change, Fig.~\ref{ellestats_width} shows the width of the distribution of $\log_{10}({\rm Pe}^{\rm loc})$.
Interestingly, the distribution is widest near the point $v_0 \simeq 10$, which is roughly where $\alpha_x > 1$.
The skewness $S$ of the distribution is also shown; $|S|$ is also enhanced in the transition regime around $v_0 \simeq 10$.
Thus, the solute distribution over pore channels with different Pe$^{\rm loc}$ values is broader and more asymmetric in the transition regime.

\begin{figure}\centering
  \includegraphics[trim=0 0 0 0,clip=,width=12cm]{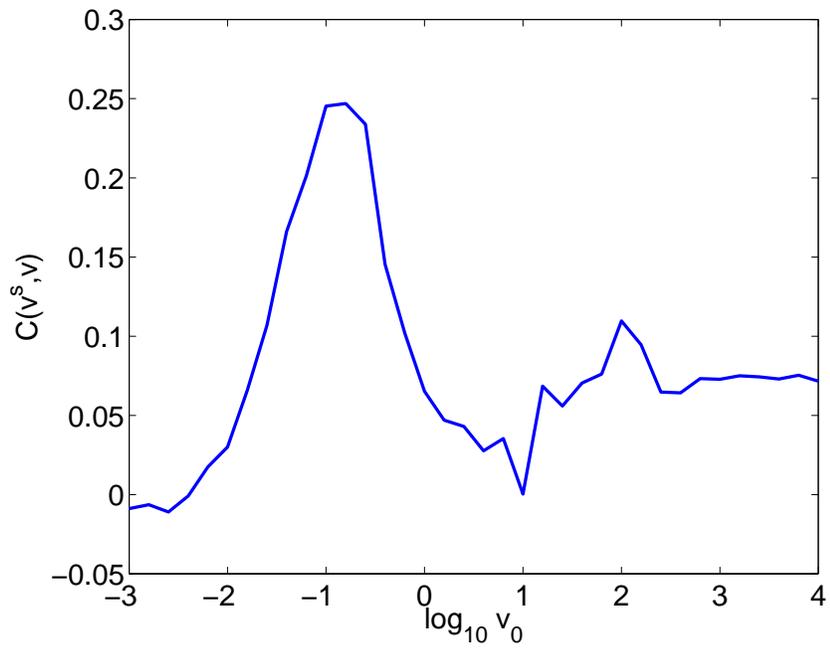}
\caption{\label{elle_corr} ``Elle'' grid. Number of nodes: 31776.
The velocity correlation defined by Eq.~(\ref{correlation}).
}
\end{figure}

The correlation between the `solute velocity' $v^s_{ij}$ and the imposed flow $v_{ij}$, calculated according to Eq.~(\ref{correlation}), is shown in Fig.~\ref{elle_corr}.
Naturally, this correlation is very small (or zero) for very low $v_0$, and positive and finite for very large $v_0$, approaching a constant value for $v_0 \gtrsim 100$.
However, there is a perhaps unexpected and rather strong correlation peak {\it prior to} the intermediate regime, around $v_0 \simeq 0.1$.
The explanation for this can be gleaned from Fig.~\ref{square_solution}. 
Although this figure refers to the square grid, the essential behavior here is similar.
At low $v_0 < 0.1$ (the `diffusive regime'), solute transport is purely diffusive and hence correlation is low. 
As $v_0$ is increased but still small (of the order of $v_0 \simeq 0.1$, the `transition regime'), the solute cloud still predominantly expands diffusively, thus reaching a large proportion of the grid, yet it starts to `sense' the advective drift in all channels, which leads to a large correlation.
At still higher $v_0$, the solute cloud is less able to expand, as it becomes confined to the predominantly advective channels, and the correlation gradually drops towards a final limiting value (corresponding to the mechanical dispersion regime, $v_0 \gtrsim 100$).
Note the oscillation between $v_0=1$ and $100$, associated with the `major regime'.

\clearpage
\section{Discussion}

Transport in the highly simplified model for porous media studied here (a network of nodes connected by one-dimensional channels, while transport through the channels is purely advective and diffusive) exhibits several of the features also observed in laboratory experiments on porous media: (a) the diffusive regime at low P\'eclet number and (b) the mechanical dispersion regime at high P\'eclet number are correctly recovered. In the intermediate regime, (c) the longitudinal dispersion generally grows faster than linear ($\alpha_x > 1$), whereas (d) the transverse dispersion grows much slower (and $\alpha_y < 1$).
An exception to this, among the various networks studied, is the tile grid, in which longitudinal connectivity was purposely reduced by a factor of two with respect to the square grid. This reduced longitudinal connectivity brings longitudinal dispersion down to a level comparable to the transverse dispersion, showing that network topology affects dispersion in a profound manner.

Fig.~\ref{square_solution} is illustrative of the generic behavior of dispersion as the flow velocity $v_0$ is increased at constant diffusivity $D$: the solute tends to be drawn preferentially into advection-dominated channels as the fluid velocity increases.
In other words, at low velocity, the solute expands randomly through all types of channel, but at high velocity, it explores much more advection-dominated (i.e., horizontal) channels than diffusion-dominated (vertical) channels.
Hence, on average, the solute will tend to spend more time in advection-dominated channels as $v_0$ is raised.
Note that this is the basic assumption underlying the heuristic model of~\cite{Milligen:2012c}.

The fact that this model qualitatively reproduces the dispersional behavior observed in actual experiments on homogeneous porous media, while Taylor dispersion is absent by definition, implies that Taylor dispersion is not required to obtain $\alpha_x > 1$ in the intermediate regime.
While this does not prove that Taylor dispersion is not relevant in actual porous media, it suggests that the relevance of Taylor dispersion for the explanation of the observed behavior ($\alpha_x>1$) needs to be reviewed carefully.

Perfectly regular grids such as the square and hexagonal grids presented above are somewhat pathological and suffer from the fact that many channels are exactly perpendicular to the mean flow direction (the $x$-axis). This means that these channels are `stagnant', and transport through them is purely diffusive, regardless of $v_0$. As $v_0$ is increased, the transport flux through these channels becomes negligible with respect to the total flux, so that the transport topology of the system becomes effectively one-dimensional. To avoid this topological breakdown, the positions of the main nodes of the network are `wiggled' slightly from their nominal positions.
With the square and hexagonal grids, the quantifier $n_v/n_0$ stagnates in the range $10 < {\rm Pe} < 100$ (cf.~Figs.~\ref{square} and \ref{hexagonal}).
It turns out that in this range, there is a gradual shift of the solute concentration towards channels with higher velocities, which however is undetected by the $n_v/n_0$ quantifier since the the corresponding part of the solute already resides in channels with Pe$^{\rm loc}>1$ (as can be verified by computing a graph like Fig.~\ref{ellestats_2d} for these cases). This effect arises because of the regular structure of the grid, causing the channels to fall into a very reduced number of classes.

This situation is considered somewhat unrealistic, as actual porous media will never be perfectly regular (not to mention the fact that there, structure is three-dimensional rather than two-dimensional).
By comparison with the regular grids, the ``Elle'' grid has the advantage of offering more gradual (realistic?) variations of quantities such as $n_v/n_0$, facilitating their study and understanding.
Thus, we focus attention on the ``Elle'' grid results.

The statistical quantifier $n_v/n_0$, indicating the `number of tracers' (amount of solute) in advective channels versus diffusive channels, increases systematically as $v_0$ is raised. The point where $n_v/n_0$ reaches 1 was found to immediately precede the moment when $\alpha_x$ reaches 1 for all grids (with the exception of the tile grid, in which $\alpha_x \le 1$).
This observation is consistent with the analytic model of~\cite{Milligen:2012c}, in which the growth of the quantifier $n_v/n_0$ with $v_0$ was hypothesized to be the cause of $\alpha_x>1$ in the intermediate P\'eclet number regime in porous media: as $v_0$ increases, not only does the velocity of the solute tracers increase (trivially), but also the number of tracers in advective ($n_v$) versus diffusive ($n_0$) channels, thus giving rise to a faster than linear increase of global longitudinal dispersion in the intermediate regime.

Furthermore, with the ``Elle'' (two-dimensional foam) network, Fig.~\ref{elle} shows that $n_v/n_0$ grows linearly, $n_v/n_0 \propto v_0^\gamma$, $\gamma \simeq 1.0$, in the range $6 < v_0 < 200$.
For smaller values of $v_0$, fluid flow is so slow that no channels are classified as `advective' ($n_v \simeq 0$), while for higher values of $v_0$, no channels are classified as `diffusive' ($n_0\simeq 0$). Thus, the fact that this power law only appears to exist over a finite range of $v_0$ values is merely due to the crudeness of the channel classification criterion. The main message of this result is that the ratio of solute in `advective'  and `diffusive' channels increases systematically as the fluid flow $v_0$ is increased, in accordance with the prediction of~\cite{Milligen:2012c}.

When plotting the distribution of solute over channels with different Pe$^{\rm loc}$ values as a function of the fluid velocity $v_0$ (Fig.~\ref{ellestats_2d}), it appears that the system experiences a `phase transition' from low fluid velocities ($v_0 \lesssim 1$) to high velocities ($v_0 \gtrsim 10$). This phase transition is reflected in the width and skewness of the (logarithmic) distribution of Pe$^{\rm loc}$, Fig.~\ref{ellestats_width}, which peak at the transition. The phase transition arises spontaneously from the complexity of the flow patterns in the two-dimensional network, and is related to the super-linear growth of the dispersion $D_x({\rm Pe})$, i.e., $\alpha_x>1$. 

The idea of a phase transition is justified from the point of view that at very low velocities, 
mean solute motion is essentially independent from the imposed advection, whereas at high velocities it moves almost exclusively with the (deterministic, externally imposed) fluid flow, corresponding to a change from an uncorrelated to a correlated state.
This change of state is not gradual, but involves a rather sharp correlation peak, as shown in Fig.~\ref{elle_corr}, which can be understood from the explanation given in the previous section.
Here, we simply note that a sharp increase of correlation is typical of phase transitions~\cite{Stanley:1971}.

Note that the quantifier $n_v/n_0$ plays the role of an `order parameter' of the phase transition. 
In particular, its derivative with respect to Pe (or `susceptibility') is large at around $v_0=1-5$ (Fig.~\ref{elle}), which roughly corresponds to the point where the width of the distribution of Pe$^{\rm loc}$, Fig.~\ref{ellestats_width}, becomes large.

We emphasize that this model only offers a limited degree of realism due to its simplicity. 
However, the statistical analysis outlined here should be easy to apply to more realistic simulations of tracer transport in porous media~\cite{Mostaghimi:2010,Ovaysi:2011}
and to detailed tracer velocity measurements in actual porous material samples facilitated by NMR techniques~\cite{Gladden:2011}.
That would provide final and definitive support for the clarification of the origin of the enhanced dispersion exponent, $\alpha_x>1$, in the intermediate regime, as discussed here.
Also, the analysis of the motion of individual tracers should allow a more complete analysis of the phase transition indicated here: tracer motion makes a transition from random motion (in the diffusive phase) to deterministic motion (in the advective phase). This change should be reflected in the correlation length between the velocities of individual tracers.

\clearpage
\section{Conclusions}

In this work, we have modeled transport through porous media by means of a two-dimensional network of one-dimensional channels (`pores') linking nodes. Along each channel, advection and diffusion was modeled using the one-dimensional advection-diffusion equation. The numerical solution method is based on the requirement that no solute is lost between network nodes. This fact stabilizes the numerical time evolution of the solution, even at very high advective velocities. The latter constitutes an advantage for the study of dispersion in a network with a wide range of local flow velocities.

We studied a range of network types, advancing the time solution from an initial state in which the solute was concentrated at a central node to a time when a stop criterion was satisfied. At the stop time, the effective longitudinal and transverse dispersion coefficients were determined. For each network, a range of flow velocities was explored.

The results indicate that the main effects of dispersion in porous media observed in laboratory experiments in homogenous  porous media (in particular, the behavior of the dispersion coefficients in the intermediate regime, as already predicted in~\cite{Milligen:2012c}) can be reproduced using a minimal model containing only the mentioned ingredients: a simplified pore geometry (network), diffusion, and advection.
Thus, some hypothetic dispersion mechanisms invoked by other authors (e.g., Taylor dispersion, anomalous diffusion or fractality~\cite{Sahimi:1993,Berkowitz:1997,Meerschaert:2001,Levy:2003}) are not needed to obtain these effects, as the observed dispersional behavior already emerges spontaneously in this highly simplified model.

The results also constitute a verification of the basic assumption underlying the analytic model of~\cite{Milligen:2012c}, namely the growth of $n_v/n_0$ as $v_0$ is increased, although the actual functional form depends on the specific network type.
This novel quantity therefore plays a crucial role in the understanding of the origin of the nonlinear behavior of the dispersion coefficients in the intermediate regime.

An analysis of the distribution of solute over channels with specific Pe$^{\rm loc}$ values indicates that this distribution experiences a `phase transition' from an unordered low-velocity state to an ordered high-velocity state. The width and skewness of the distribution peak in the intermediate regime, at the velocity ($v_0 \simeq 10$) where the growth exponent of longitudinal dispersion ($\alpha_x$) exceeds 1.

\section*{Acknowledgements}
Research sponsored in part by the Ministerio de Econom\'ia y Competitividad of Spain under project Nr.~ENE2012-30832.
This study was carried out within the framework of DGMK (German Society for Petroleum and Coal Science and Technology) research project 718 ``Mineral Vein Dynamics Modelling'', which is funded by the companies ExxonMobil Production Deutschland GmbH, GDF SUEZ E\&P Deutschland GmbH, RWE Dea AG and Wintershall Holding GmbH, within the basic research program of the WEG Wirtschaftsverband Erd\"ol- und Erdgasgewinnung e.V. We thank the companies for their financial support and their permission to publish these results. 



\end{document}